\documentclass{du-journals}
\journal{Iranian Journal of Astronomy and Astrophysics}

\title{Solar Mini-Dimming Kinematics and Their Positive Correlations with  Coronal Mass Ejections  and Prominence}
\subtitle{Solar Mini-Dimming   Kinematics ...}

\author[1]{Nasibe Alipour}
\author[2]{Hossein Safari}
\address[1]{Department of Physics, University of Zanjan, Zanjan, Iran}

\begin{document}

\begin{abstract}
Solar mini-dimmings can be detect in the Extreme Ultra-Violet
 coronal eruptions. Here, sequences of 171\AA\ images taken
by Solar Dynamic Observatory/Atmospheric Imaging Assembaly
on 13 June 2010 are used. In this special day,
both of coronal mass ejection and prominence were observed.
 The average velocities and accelerations  of 500
mini-dimmings which were detected using on feature based
classifier (Alipour et al 2012) are studied. The large number of
mini-dimmings shows positive accelerations in the beginning times
as similar as Coronal Mass Ejections.
 On the the start time of  prominence eruptions, the number of mini-dimmings is increased to
a maximum values. There is a positive correlation between the
kinematics of  mini-dimmings and both CME and prominences. This
study can be extended to understand the exact relationship of
 CMEs and  mini-dimmings.
\end{abstract}

\section{Introduction}
Corona mass ejections (CMEs) are bursts of clods of magnetic field
and plasma from the solar corona into the interplanetary space. In
coronagraph images a classical CME has often
 a three-part structure consists of a bright outward,
 dark cavity behind and a bright dense
 core inward \cite{Lang}. Corona mass ejections are often associated
 with forms of solar activity
 that the most important of these are solar flare,
  eruptive prominence, coronal dimming
 and EIT wave. Solar flares are sudden, vigorous outbursts and
 compendium of ionized and hot gas
close to sunspots. Protons and electrons accelerate into magnetic
loops and release
 energy equivalent to the millions hydrogen bombs in a short period of time.
  Solar flares are
 classified according to their strength and are known  in five
 classes A, B, C, M, X  (from weak to strong). About 60\%
 of flares (M and X classes) are associated with CMEs and about
 50\% of CMEs are accompanied with flares. Solar prominence are
 dense clouds of material similar to Phoenix suspended above the solar surface.
 The dense core of a CME structure is an erupting prominence.
 In coronagraph
 images, erupting prominence are observed as a bright area. About of 75\% of CMEs are
 associated with prominence \cite{Subramanian}. Coronal dimmings are similar to transitory coronal
 holes and are usually observe as density depletion in both X-rays and Extreme UltraViolet
 images. The dimmings are formed due to the depletioned plasmas by the eruption of the local
 magnetic field and in some cases it due to the temperature gradient \cite{Thompson}.
 Wave-like brightening propagate following of dimming expansion (approximately
 circular) over the whole of the Sun that are known as EIT wave \cite{Podladchikova}.
  This group
 of waves propagate from the active region and continue to where
 the magnetic field lines are further \cite{Chen}.

Mini-dimmings were first studied by Innes et al. (2009) by using
 171 $\AA$ space-time slices \cite{Innes}. Alipour, Safari, and
Innes (2012), extended a method of automatic detection of
mini-dimming using invariant properties of Zernike moments and
feature based classifier Support Vector Machine (SVM)
\cite{Alipour}. Their method detected the groups of small dimming
features and the sizes, time durations, and velocities
distributions were studied. The relation between corona mass
ejections and mini-dimmings
 are studied by Alipour \& Safari (2012) \cite{Alipour2}.

 Here,
 in similar studies (Alipour et al. and Alipour \& Safari), we used Solar Dynamic
Observatory/Atomospheric Imaging Assembly (SDO/AIA) images to
investigate some properties of mini-dimming (average velocity and
acceleration). The relation between mini-dimmings and  prominence
eruptions are presented. The paper layout is as follows:  the data
analysis is discussed in
 Section \ref{sect2}. The  average
 velocities and accelerations of mini-dimmings are studied  in Section \ref{sect3}.
  Statistical studies on the
 relation between prominence and mini-dimmings are given in
 Section \ref{sect4}. The result and conclusion are presented in
 Section \ref{sect5}.
\section{Data Analysis}\label{sect2}
Both of corona mass ejections and a prominence were occured
 on 13 June 2010, simultaneously. This prominence was observed at
304 \AA\ (50000 K), and CMEs are taken from SOHO/LASCO coronagraph
 ({http://cdaw.gsfc.nasa.gov/CME list}).

\begin{center}
\begin{figure}
\includegraphics{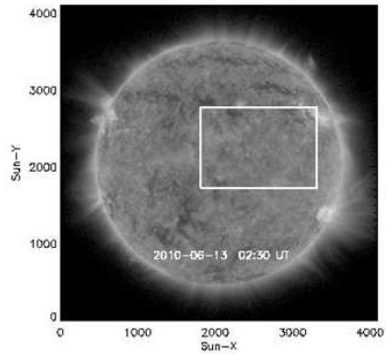}
      \vspace{4.8cm}
          \caption[]{ The solar full disk in 171\AA wavelength at 00:48 UT on 13 June 2010 by SDO/AIA. The white box delineates area
 of 1400$\times$1050 pixels for determine the number of mini-dimmings.}
\label{fig1}
   \end{figure}
   \end{center}

We used 171 $\AA$  (0.6 MK) SDO/AIA images (1400$\times$1050
pixels)
 with a time cadence of 2.5 minutes and a pixel size
    0.6 arcsec, taken on 13 June 2010 from cut out services. See
Figure \ref{fig1}. Following Alipour et al. (2012), an automatic
mini-dimming detection procedure applied
 to systematically scans through space-time slices of the data
(see flowchart of their algorithm, Figure 3 therein). The number
of 350 space-time slices with the size of 1400$\times574$ are
analyzed.

\begin{figure} 
 \centerline{\includegraphics[width=10cm]{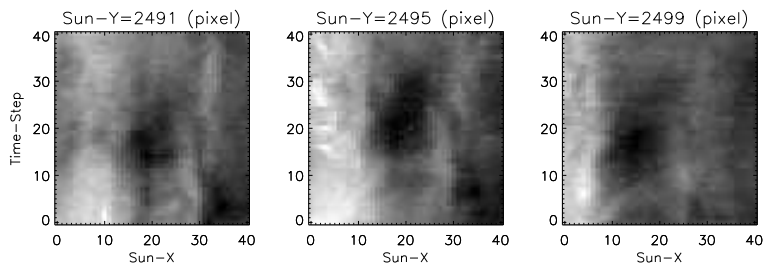}}
 \caption[]{The spatial expansion of a mini-dimming along the Sun-y.}
 \label{fig2}
\end{figure}

\section{Acceleration and velocity of mini-dimmings}\label{sect3}
\begin{figure} 
 \centerline{\includegraphics[width=15cm]{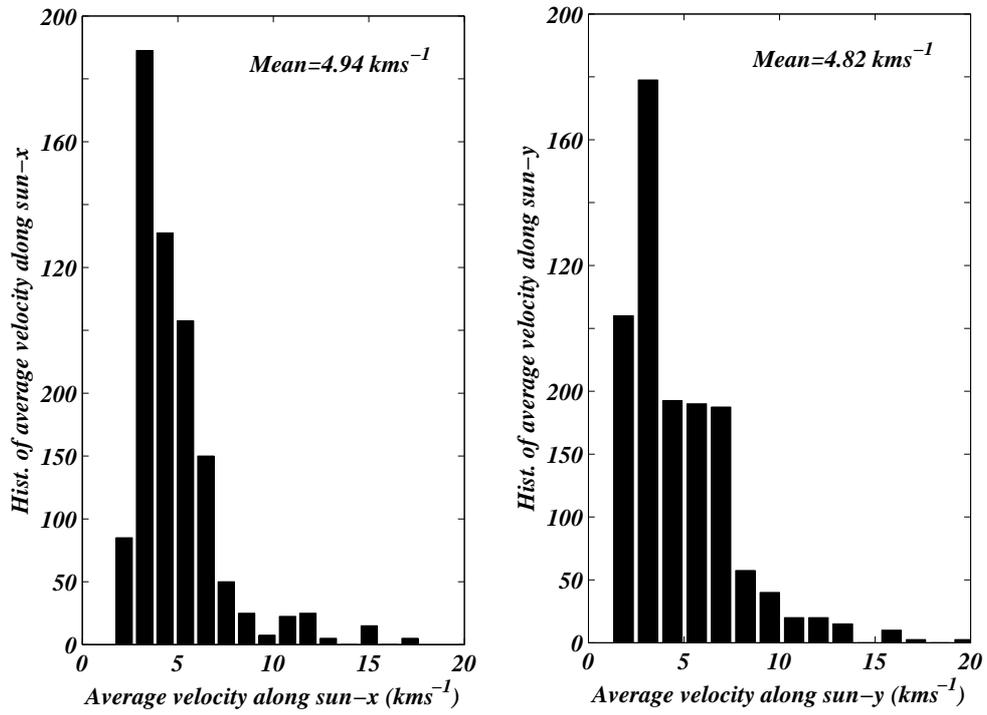}}
 \caption[]{The histogram of average velocity of 500 mini-dimmings along sun-x (left) and sun-y (right) directions in SDO/AIA
  images on 13 June 2010 plotted versus time.}
 \label{fig3}
\end{figure}

\begin{figure} 
 \centerline{\includegraphics[width=10cm]{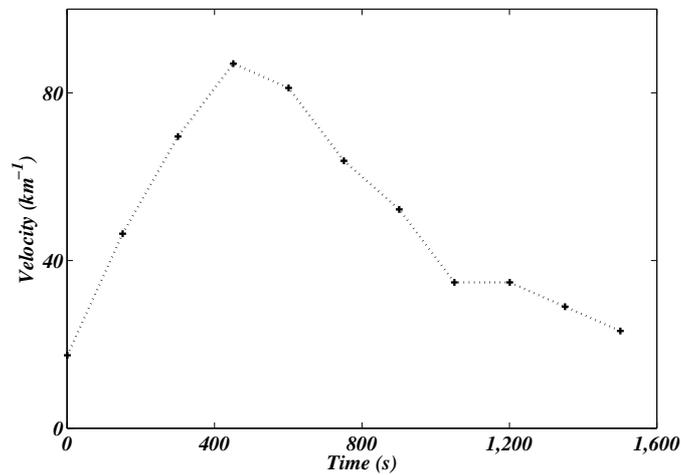}}
 \caption[]{The velocity of a mini-dimming is plotted versus time.}
 \label{fig4}
\end{figure}

\begin{figure} 
 \centerline{\includegraphics[width=10cm]{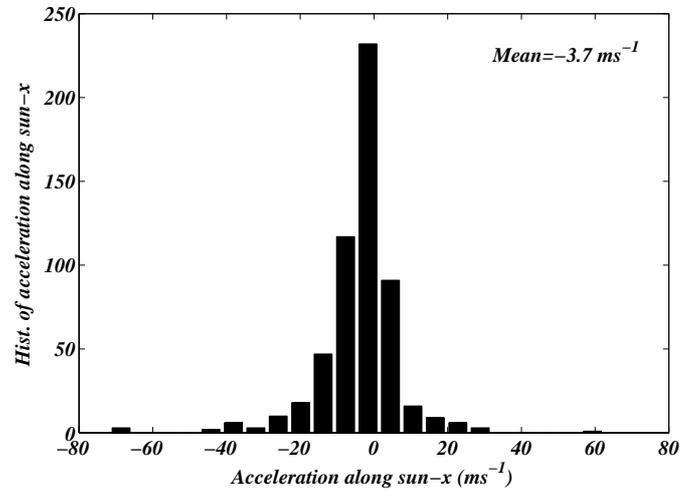}}
 \caption[]{The histogram of acceleration of 500 mini-dimmings are plotted.}
 \label{fig5}
\end{figure}

The average velocities and accelerations of detected mini-dimmings
are computed. To do this, all events with spatial expansions
($\Delta x < 20$ arcsec and $\Delta y < 20$ arcsec) and temporal
expansions ($\Delta t < 20$ min) are grouped, Alipour et al.
(2012). The center of events (position and time) of detected
mini-dimming is determined. The spatial expansion of an event
along the Sun-y is shown in Figure \ref{fig2}.  The region growing
algorithm is applied to compute the dimmings sizes (both $\Delta
x$ and $\Delta y$) and time intervals ($\Delta t$) from space-
time blocks of each group. The velocities along the Sun-x ($\Delta
x/\Delta t$) and Sun-y  ($\Delta y/\Delta t$) directions are
calculated. The velocity-time graph of an event along the Sun-x is
shown in Figure \ref{fig4}. The histograms of average velocities
for 500 mini-dimming are shown in Figure \ref{fig3}. Using
velocity-time graphs and fitting $\Delta x=\frac{1}{2}a\Delta t^2$
the accelerations, $a$, are computed for each of grouped
mini-dimmings. As shown in Figure \ref{fig4}, the velocity
increases to a maximum values with a positive acceleration and
then decreases with a negative acceleration. Mittal \& Narian
(2009) studied the distribution of speed of corona mass ejections
 from solar cycle 23 \cite{mittal}. They shown that the
 accelerations
 of corona mass ejections are positive values in the beginning.
In Figure \ref{fig5}, the histogram of accelerations along Sun-x
is presented. We see that large number of mini-dimming shows
positive acceleration.
\begin{figure} 
 \centerline{\includegraphics[width=13cm]{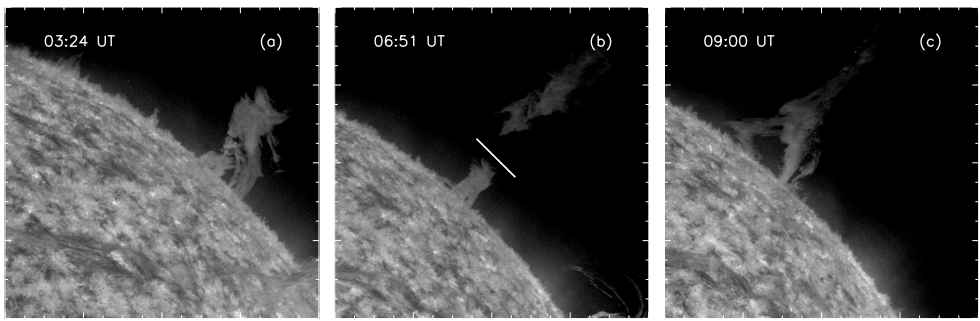}}
 \caption[]{A prominence eruption at 304 $\AA$ on 13 June 2010 (a) at
first phase 03:24 UT the proximate condition of the eruption
occurred, (b) at second phase 06:51 UT the end of eruption (the
white line indicate the different parts of plasma consists of the
eruption), (c) at third phase 09:00 UT reconnection of the
magnetic filed and the plasma ejected into the higher part of the
corona \cite{Regnier}.}
 \label{fig6}
\end{figure}
\begin{figure} 
 \centerline{\includegraphics[width=10cm]{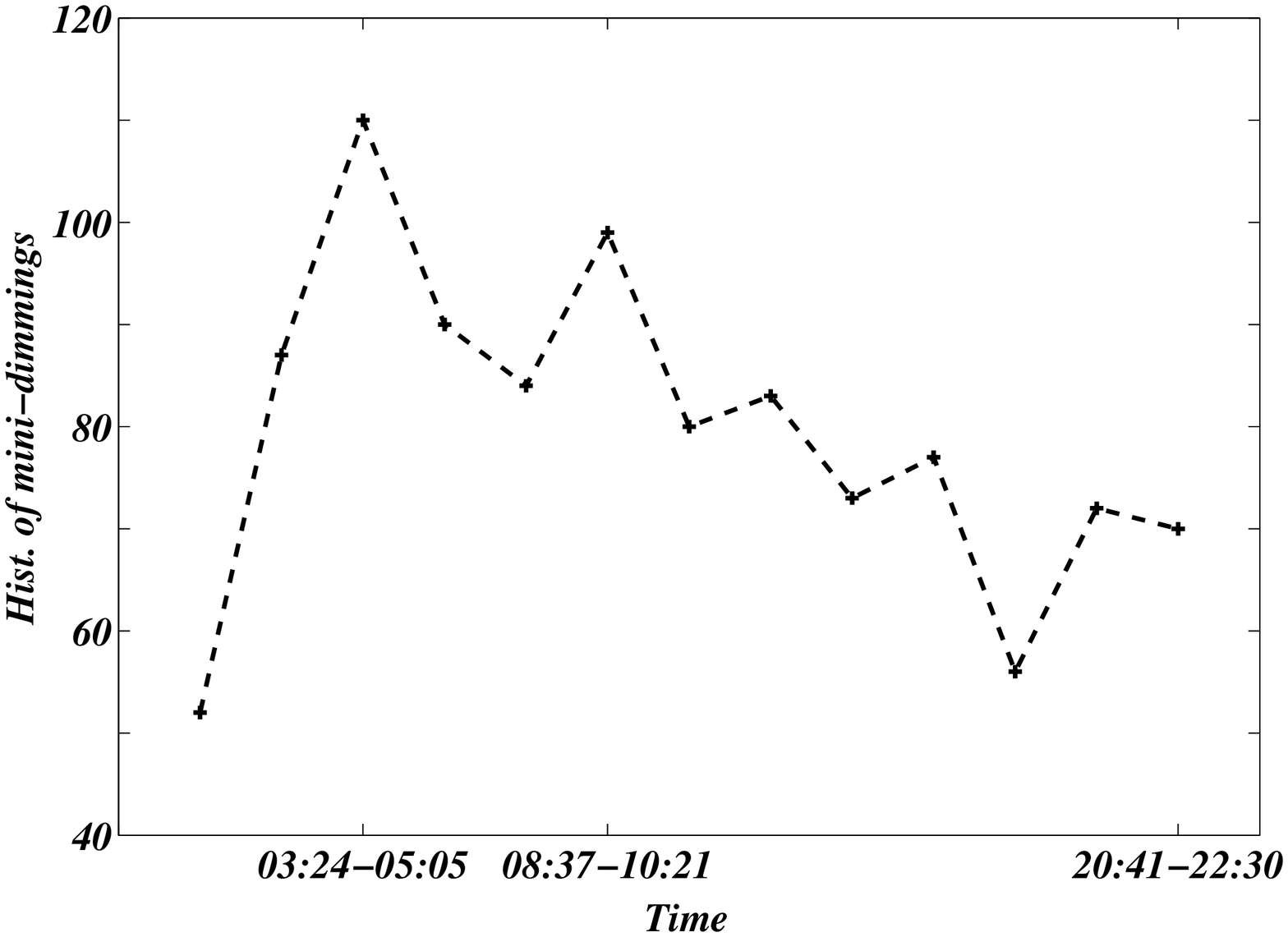}}
 \caption[]{The histogram of 500 mini-dimmings are plotted versus time. }
 \label{fig7}
\end{figure}

\section{Prominence eruption and mini-dimmings }\label{sect4}
Solar prominences are bright loops of ionized and hot gas from the
chromosphere into the corona. A single erupting prominence
involves hundreds of millions of tons from the solar particles. A
prominence eruption leading to a propagating corona mass ejection
consists of three steps are presented in Figure \ref{fig6}. In
Figure \ref{fig7}, the histogram of mini-dimmings are plotted
versus time (with interval of 100 minute). We see that, on the the
start time of eruption (first phase 03:24 UT - 05:08 UT), the
number of mini-dimmings is increased to maximum value. On the
second phase of prominence eruption (06:51 UT - 08:35 UT), the
number of mini-dimmings is decreased. In the time of reconnection
and rearrangement of magnetic field (third phase 08:37 UT - 10:21
UT), this number is increased.

\section{Results and Conclusion}\label{sect5}
Prediction of coronal mass ejections and flares are important for
solar physicist \cite{Tajfirouze}.
 Observations of eruptive dimmings and EIT waves
could be processed to predict solar CMEs, that is faster than
coronagraph observations \cite{podladchikova2012}.

 The automatic
detection method for solar mini-dimmings (Alipour et al. 2012) has
been applied to SDO/AIA 171\AA\ data. On 13 June 2010, the average
velocities and acceleration of 500 events were analyzed. We see
that, approximately the large number of mini-dimmings were
occurred with positive average velocities and accelerations at
their beginning times.  The number of detected mini-dimming
increased in the time of first phase of prominence eruption and
decrease in second phase of eruption, and again increased in the
third phase of reconnection of magnetic field. These are positive
correlations between mini-dimmings kinematics and both coronal
mass ejections and prominence. These positive relationships might
help to predict the occurrence of an associated CME.

%

\end{document}